\documentstyle[12pt,epsf]{article}
\catcode`\@=11
 
\@addtoreset{equation}{section}
\def\theequation{\arabic{equation}}

\def\@normalsize{\@setsize\normalsize{15pt}\xiipt\@xiipt
\abovedisplayskip 14pt plus3pt minus3pt%
\belowdisplayskip \abovedisplayskip
\abovedisplayshortskip  \z@ plus3pt%
\belowdisplayshortskip  7pt plus3.5pt minus0pt}
\def\small{\@setsize\small{13.6pt}\xipt\@xipt
\abovedisplayskip 13pt plus3pt minus3pt%
\belowdisplayskip \abovedisplayskip
\abovedisplayshortskip  \z@ plus3pt%
\belowdisplayshortskip  7pt plus3.5pt minus0pt
\def\@listi{\parsep 4.5pt plus 2pt minus 1pt
            \itemsep \parsep
            \topsep 9pt plus 3pt minus 3pt}}
 
\def\underline#1{\relax\ifmmode\@@underline#1\else
        $\@@underline{\hbox{#1}}$\relax\fi}
\@twosidetrue
\relax
\catcode`@=12
 
\catcode`\@=11

\def\section{\@startsection{section}{1}{\z@}{3.5ex plus 1ex minus
   .2ex}{2.3ex plus .2ex}{\large\bf}}

\def\ps@headings{\def\@oddfoot{}\def\@evenfoot{}
\def\@oddhead{\hbox{}\hfill
        \makebox[.5\textwidth]{\raggedright\ignorespaces --\thepage{}--
        \hfill }}
\def\@evenhead{\@oddhead}
\def\subsectionmark##1{\markboth{##1}{}}
}
 
\ps@headings
 
\catcode`\@=12
 
\relax

\def\figcap{\section*{Figure Captions\markboth
        {FIGURECAPTIONS}{FIGURECAPTIONS}}\list
        {Fig. \arabic{enumi}:\hfill}{\settowidth\labelwidth{Fig. 999:}
        \leftmargin\labelwidth
        \advance\leftmargin\labelsep\usecounter{enumi}}}
 \relax
\def\tablecap{\section*{Table Captions\markboth
        {TABLECAPTIONS}{TABLECAPTIONS}}\list
        {Table \arabic{enumi}:\hfill}{\settowidth\labelwidth{Table 999:}
        \leftmargin\labelwidth
        \advance\leftmargin\labelsep\usecounter{enumi}}}
 \relax
\def\reflist{\section*{References\markboth
        {REFLIST}{REFLIST}}\list
        {[\arabic{enumi}]\hfill}{\settowidth\labelwidth{[999]}
        \leftmargin\labelwidth
        \advance\leftmargin\labelsep\usecounter{enumi}}}
 \relax
 
\catcode`\@=11
 
\def\marginnote#1{}
\newcount\hour
\newcount\minute
\newtoks\amorpm
\hour=\time\divide\hour by60
\minute=\time{\multiply\hour by60 \global\advance\minute by-
\hour}
\edef\standardtime{{\ifnum\hour<12 \global\amorpm={am}%
    \else\global\amorpm={pm}\advance\hour by-12 \fi
    \ifnum\hour=0 \hour=12 \fi
    \number\hour:\ifnum\minute<100\fi\number\minute\the\amorpm}}
\edef\militarytime{\number\hour:\ifnum\minute<100\fi\number\minute}
\def\draftlabel#1{{\@bsphack\if@filesw {\let\thepage\relax
  \xdef\@gtempa{\write\@auxout{\string
    \newlabel{#1}{{\@currentlabel}{\thepage}}}}}\@gtempa
    \if@nobreak \ifvmode\nobreak\fi\fi\fi\@esphack}
     \gdef\@eqnlabel{#1}}
\def\@eqnlabel{}
\def\@vacuum{}
\def\draftmarginnote#1{\marginpar{\raggedright\scriptsize\tt#1}}
\def\draft{\oddsidemargin -.5truein
        \def\@oddfoot{\sl preliminary draft \hfil
        \rm\thepage\hfil\sl\today\quad\militarytime}
        \let\@evenfoot\@oddfoot \overfullrule 3pt
        \let\label=\draftlabel
        \let\marginnote=\draftmarginnote
   
\def\@eqnnum{(\theequation)\rlap{\kern\marginparsep\tt\@eqnlabel}%
\global\let\@eqnlabel\@vacuum}  }
\def\preprint{\twocolumn\sloppy\flushbottom\parindent 1em
        \leftmargini 2em\leftmarginv .5em\leftmarginvi .5em
        \oddsidemargin -.5in    \evensidemargin -.5in
        \columnsep 15mm \footheight 0pt
        \textwidth 250mmin      \topmargin  -.4in
        \headheight 12pt \topskip .4in
        \textheight 175mm
        \footskip 0pt
\def\@oddhead{\thepage\hfil\addtocounter{page}{1}\thepage}
        \let\@evenhead\@oddhead \def\@oddfoot{} \def\@evenfoot{} 
}
\def\titlepage{\@restonecolfalse\if@twocolumn\@restonecoltrue\onecolumn
     \else \newpage \fi \thispagestyle{empty}\c@page\z@
        \def\thefootnote{\fnsymbol{footnote}} }
\def\endtitlepage{\if@restonecol\twocolumn \else  \fi
        \def\thefootnote{\arabic{footnote}}
        \setcounter{footnote}{0}}  %\c@footnote\z@ }
\catcode`@=12
\relax
\def\ps@headings{\def\@oddfoot{}\def\@evenfoot{}
\def\@oddhead{\hbox{}\hfill
        \makebox[.5\textwidth]{\raggedright\ignorespaces --\thepage{}--
        \hfill }}
\def\@evenhead{\@oddhead}
\def\subsectionmark##1{\markboth{##1}{}}
}
 
\ps@headings
 
\relax

\def\firstpage#1#2#3#4#5#6{
\begin{document}
%\draft
%%%%%%%%%%%%%%%%% MACROS %%%%%%%%%%%%%%%%%%
\def\beq{\begin{equation}} 
\def\eeq{\end{equation}} 
\def\bea{\begin{eqnarray}} 
\def\eea{\end{eqnarray}} 
\def\bq{\begin{quote}} 
\def\eq{\end{quote}}
\def\ra{\rightarrow} 
\def\lra{\leftrightarrow} 
\def\ups{\upsilon}
\def\bq{\begin{quote}} 
\def\eq{\end{quote}}
\def\ra{\rightarrow} 
\def\un{\underline}
\def\ov{\overline}
\newcommand{\cm}{Commun.\ Math.\ Phys.~}
\newcommand{\prl}{Phys.\ Rev.\ Lett.~}
\newcommand{\pr}{Phys.\ Rev.\ D~}
\newcommand{\pl}{Phys.\ Lett.\ B~}
\newcommand{\ibar}{\bar{\imath}}
\newcommand{\jbar}{\bar{\jmath}}
\newcommand{\np}{Nucl.\ Phys.\ B~}
\newcommand{\F}{{\cal F}}
\renewcommand{\L}{{\cal L}}
\newcommand{\A}{{\cal A}}
\def\154{\frac{15}{4}}
\def\153{\frac{15}{3}}
\def\32{\frac{3}{2}}
\def\254{\frac{25}{4}}
\begin{titlepage}
\nopagebreak
\title{\begin{flushright}
        \vspace*{-1.8in}
        {\normalsize hep-ph/9704336}\\[-9mm]
        {\normalsize SHEP 97/04}\\[-9mm]
        {\normalsize IOA-TH/97-005}\\[3mm]
\end{flushright}
\vfill
{#3}}
\author{\large #4 \\[1.0cm] #5}
\maketitle
\vskip -7mm     
\nopagebreak 
\begin{abstract}
{\noindent #6}
\end{abstract}
\vfill
\begin{flushleft}
\rule{16.1cm}{0.2mm}\\[-3mm]
$^{\star}${\small Research supported in part by grant numbers
GR/K55738, TMR contract ERBFMRX-CT96-0090 and $\Pi$ENE$\Delta$-15815/95}\\ 
April 1997
\end{flushleft}
\thispagestyle{empty}
\end{titlepage}}
 
\def\simlt{\stackrel{<}{{}_\sim}}
\def\simgt{\stackrel{>}{{}_\sim}}
\date{}
\firstpage{3118}{IC/95/34}
{\large\bf  Leptoquarks in SUSY Unified Models 
and the HERA Events$^{\star}$} 
{S. F. King$^{\,a}$  and G.K. Leontaris$^{\,b}$}%\\[-3mm]
{\normalsize\sl
$^a$Department of Physics,University of Southampton,
{}Southampton,  SO17 1BJ,  UK\\[-3mm]
\normalsize\sl
$^b$ Theoretical Physics Division, Ioannina University,
GR-45110 Ioannina, Greece.}
{Motivated by the recent HERA events which are consistent
with a possible leptoquark interpretation, we discuss the prospects
for including additional light colour triplets and anti-triplets
in the spectrum of supersymmetric
unified theories. We focus on a particular string-inspired Pati-Salam
model, and propose a simple mechanism by which
a light colour triplet of charge -1/3 plus anti-triplet of charge 1/3,
may have a mass of order 200 GeV, with one of the new states
having leptoquark couplings and with proton decay suppressed.
We also discuss possible scenarios for gauge unification in such a model.}
{.}

\newpage

The recent HERA events \cite{H1,Zeus} have been followed by much
theoretical speculation about leptoquarks
\cite{leptoquarks}.
If we accept that the excess of events reported  at high - $Q^2$ 
is not a statistical fluctuation and future runs confirm their
existence, it is clear that they are suggestive of new physics 
beyond the standard model (SM). In particular, assuming that the
events occur at the s-channel, then the observed peak at definite large
$x$-values is a distribution which corresponds to a mass determination
of order ${\cal O} (200)$ GeV.  The various theoretical explanations 
include  contact interactions, $R$- parity violation and leptoquarks.  
In this letter, prompted by the above experimental findings at
HERA,  we discuss the prospects for incorporating
leptoquarks into supersymmetric
unified models. \footnote{By leptoquarks we mean
light colour triplets with leptoquark couplings.
The introduction such states might be expected to destroy
the unification of the three gauge couplings, however there are various
ways to remedy this as discussed later. See also \cite{lv}. }

${\cal A)}$. 
To set the notation we first present the
R-parity conserving superpotential
in the MSSM:
\bea
{\cal W} &=& \lambda_1 q u^c h_2 +\lambda_2 q d^c h_1 +
              \lambda_3 \ell e^c h_1 +\lambda_4 \phi_0 h_1h_2
\label{w}
\eea
where $q=(u,d)=(3,2,\frac{1}{6})$, $u^c=(\bar{3},1,-\frac{2}{3})$,
$d^c=(\bar{3},1,\frac{1}{3})$, $\ell =(\nu ,e)=(1,2,-\frac{1}{2})$, 
and $e^c=(1,1,1)$ 
are the left-handed quark and lepton
superfields which transform under the standard model gauge group as shown
and $h_{1,2}$ the standard higgses. $\phi_0$ is
a singlet which realises the higgs mixing.\footnote{
This is actually the next-to-minimal supersymmetric standard model
(NMSSM). In the MSSM Higgs mixing is achieved by a
$\mu h_1 h_2$ coupling.}
In addition, one may add the following interactions
\bea
{\cal W}'&=& \lambda_5 \ell\ell e^c +\lambda_6\ell q d^c +
             \lambda_7 u^c d^c d^c + \lambda_8\phi_0h_2\ell
             \label{RvW}
 \eea
When both $ \lambda_{5,6}$ couplings are present, there
are graphs mediated by the scalar partners which lead to
lepton number violation. On the other hand, the coexistence
of $ \lambda_{6,7}$ couplings leads to fast proton decay,
unless the couplings are unnaturally small \cite{pd}. 
A natural way to avoid such problematic couplings is to impose
R-parity which forbids  all the terms in (\ref{RvW}), unless
initial conditions on the couplings 
at the unification scale are assumed 
in order to allow only the desired terms at low energy
(see for example \cite{bbs}).
Henceforth we shall assume such an R-parity, and turn instead to
the possibility of leptoquarks.

Assuming the existence of light leptoquarks the basic question we
wish to address is how they might naturally
be incorporated within the framework
of supersymmetric unified models. 
{}From this point of view leptoquarks correspond to
new coloured states which are remnants of representations
of a higher symmetry. For example in the context of 
a non-supersymmetric $SU(5)$ theory additional light 
states $Q=(3,2,\frac{1}{6})$
and $\bar{Q} = (\bar{3},2,-\frac{1}{6})$ contained in the $10 + \bar{10}$
were introduced to adjust the wrong prediction of $\sin^2\theta_W$
\cite{my}. However tremendous fine-tuning is necessary to
split apart the light $Q$ state from the remaining components of the
$10$ which must remain superheavy. This is similar to the problem
of splitting the Higgs doublet from the Higgs colour triplet
in the $5$ of $SU(5)$. From this example it would seem that light leptoquarks
only serve to exacerbate the doublet-triplet splitting problem
present in unified models, supersymmetric or not.
However, as we shall see in the next section, there is a natural way to obtain
a pair of light leptoquarks in supersymmetric unified models
without any fine-tuning.

A further candidate for a light leptoquark, common to
all grand unified models, are new left-handed
representations $D^c=(\bar{3},1,\frac 13)$ and
$\bar{D}^c=(3,1,- \frac 13)$ where $D^c$ has the quantum numbers of 
the down quark singlet $d^c$.
There are two types of couplings which can exist in the
low energy superpotential. These are,
\bea
 {\cal W}_1 &=& \lambda_9 q q \bar{D}^c +\lambda_{10} u^c d^c D^c 
\label{w1}\\
{\cal W}_2 &=& \lambda_{11} D^c q \ell + \lambda_{12} \bar{D}^c u^c e^c +
           \lambda_{13} \bar{D}^c d^c \nu^c\label{w2}
\eea
where we have assumed that $\nu^c$ is the right handed neutrino.
In order to avoid proton decay problems,
 with a suitable discrete symmetry we may 
 prevent one of ${\cal W}_1,{\cal W}_2$. There are other
exciting possibilities of exotic quark states
\cite{leptoquarks,bkmw}, which create couplings 
that might interpret the HERA data.
Thus, states like those described above, offer interesting
possibilities for new phenomenology.
Since all of these new states carry colour, the new couplings
should not lead to fast proton decay. In particular, symmetries
imposed by hand in the above superpotential pieces are not
always consistent with the unified gauge symmetry.

${\cal B})$. After these rather general considerations 
we now specialise to a particular model in which
it is possible to have light leptoquarks 
without inducing excessive proton decay,
and to achieve this in a natural way without any fine tuning.
This is the string-inspired Pati-Salam model \cite{pati,alr}.
Here we briefly summarise the parts of the model
which are relevant for our analysis. The gauge group is,
\begin{equation}   
G_{PS}=
\mbox{SU(4)}\otimes \mbox{SU(2)}_L \otimes \mbox{SU(2)}_R. \label{422}
\end{equation} 
The left-handed quarks and leptons are accommodated in 
the  representations
\bea
{F} & = & (4,2,1)=(q,l) \nonumber \\
{\bar{F}} & = & (\bar{4},1,\bar{2})= (u^c,d^c,\nu^c,e^c ) \nonumber 
\eea

The MSSM Higgs fields are contained in the following representations,
\begin{equation} h_{a}^x=(1,\bar{2},2)=
\left(\begin{array}{cc} {h_2}^+ & {h_1}^0 \\ {h_2}^0 & {h_1}^- \\
\end{array} \right) \label{h}
\end{equation} 
Under the symmetry breaking in
Eq.\ref{422to321}, the Higgs field $h$ in Eq.\ref{h} splits into two 
Higgs doublets $h_1$, $h_2$ whose neutral components subsequently    
develop weak scale VEVs,
\begin{equation} <h_1^0>=v_1, \ \ <h_2^0>=v_2 \label{vevs}
\end{equation} with $\tan \beta \equiv v_2/v_1$. The spectrum of
the model is completed with four singlets $\varphi,\phi_{i}$,
$i=1,2,3$ where $<\varphi >\sim \mu$ realises the higgs mixing
and $\phi_i$ mix with the right handed neutrinos and
participate in the higgs mechanism\cite{alr}.

The Pati-Salam gauge symmetry is broken at the scale $M_{PS}$ 
by the following Higgs representations 
\bea
{\bar{H}} & = & (\bar{4},1,\bar{2})= (u^c_H,d^c_H,\nu^c_H,e^c_H ) \nonumber \\
{H} & = & (4,1,2) = (\bar{u}^c_H,\bar{d}^c_H,\bar{\nu}^c_H,\bar{e}^c_H ) 
\eea
The neutral components of the Higgs fields are assumed to develop VEVs
\begin{equation} <\tilde{\nu_H^c}>=<{\tilde{\bar{\nu_H^c}}}>\sim M_{PS},
\label{HVEV}
\end{equation} leading to the symmetry breaking at $M_{PS}$:
\begin{equation}
G_{PS}\rightarrow
\mbox{SU(3)}_C \otimes \mbox{SU(2)}_L \otimes \mbox{U(1)}_Y
\label{422to321}
\end{equation} in the usual notation.  

The high energy Higgs mechanism removes the $H,\bar{H}$ components
$u^c_H,e^c_H,\bar{u}^c_H,\bar{e}^c_H$ from the physical spectrum
(half of these states get eaten by the heavy gauge bosons and gauginos
and the other half will become massive Higgs bosons),
leaving massless $d^c_H,\bar{d}^c_H$. In order to give these
states a large mass one introduces a colour sextet
superfield 
$$
D_6=(6,1,1)=(D^c,\bar{D^c}),
$$
where as before
$D^c=(\bar{3},1,\frac 13)$ and
$\bar{D}^c=(3,1,- \frac 13)$. 
We take the gauge invariant superpotential to have the form
(dropping all coupling constants)
\bea
{\cal W}_{422}&\sim & \bar{F}F h + \bar{F}H\phi_i +\varphi (h h +\phi_i\phi_j
+D_6D_6 + H\bar{H})\nonumber\\
          &+& F{F}D_6 +\bar{F}\bar{F}D_6+\bar{H}\bar{H}D_6 + H H D_6
          \label{W422}
          \eea
Now, the way the colour triplets receive 
superheavy masses is the following.
Remember first that the decomposition of the sextet gives an
antitriplet/triplet pair 
($D_6\ra D^c+\bar{D^c}$). On the other hand
$\bar{H}, H$ fields contain also another such pair with the
same quantum numbers: $d^c_H, \bar{d}_H^c$. 
To break the $SU(4)\times SU(2)_R$
the $H,\bar{H}$ fields acquire vevs of ${\cal O}(M_{PS})$. Then from
the terms of the second line in (\ref{W422}) one  gets the 
following two mass terms
 \beq
 <H> H D_6 +<\bar{H}>\bar{H} D_6\ra <\tilde{\bar{\nu}}_H>\bar{d}^c_HD^c+
 <\tilde\nu_H^c>d_H^c \bar{D^c}
 \label{Dmass}
 \eeq

{}From the point of view of proton decay the most dangerous colour
triplets are those contained in the
heavy sextet field $D_6$. 
This is because of
the terms in the second line of the superpotential 
which mix the families with them.
Indeed, the following dangerous combinations of couplings appear
 \bea
 F{F}D_6 &\ra& \lambda_{9}qq\bar{D^c} +\lambda_{11} q\ell D^c\\
 \bar{F}\bar{F}D_6 &\ra& \lambda_{10}u^cd^cD^c + 
\lambda_{12}u^ce^c\bar{D^c}+ \lambda_{13}d^c \bar{D^c}\nu^c
 \eea
On the right hand side of the above equations we have inserted
the same couplings used in equs(\ref{w1},\ref{w2}) in order to 
emphasise the way that GUT models in general lead to couplings
which can potentially generate proton decay.

${\cal C})$. 
The question we now ask is: can we somehow have a light
colour triplet plus anti-triplet without inducing excessive
proton decay? 
At first sight this would seem unlikely due to the 
dimension-5 operators generated by the 
colour triplet exchange diagrams in Figs.1,2.

\begin{figure}
\begin{center}
\leavevmode   
\hbox{\epsfxsize=3in
\epsfysize=2in
\epsffile{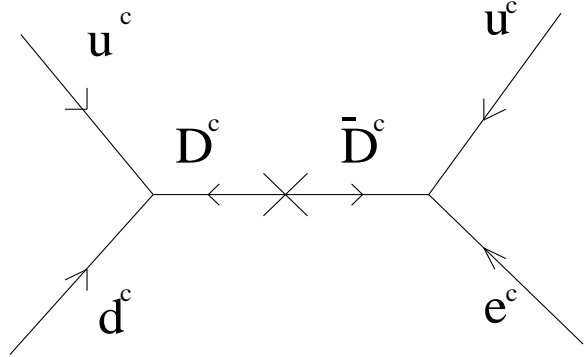}}
\end{center}
\caption{A dimension-5 proton decay operator generated from 
terms in the operator $\bar{F}\bar{F}D_6$.}
\label{Fig.1}
\end{figure}

\begin{figure}
\begin{center}
\leavevmode   
\hbox{\epsfxsize=3in
\epsfysize=2in
\epsffile{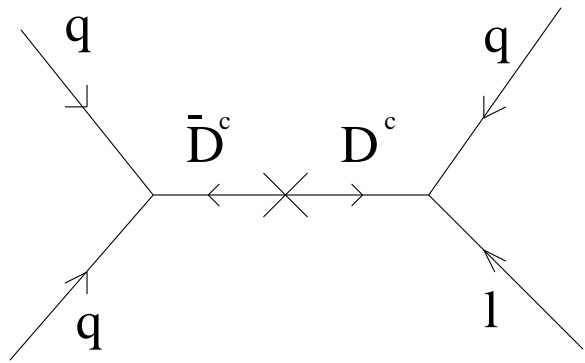}}
\end{center}
\caption{A dimension-5 proton decay operator generated from 
terms in the operator ${F}{F}D_6$.}
\label{Fig.2}
\end{figure}

Note that the dimension-5 proton decay diagrams rely on the mass mixing of the
$D^c$ and $\bar{D^c}$ colour triplets, which is controlled by the
term $\varphi D_6 D_6$ in the superpotential whose adjustable coupling
strength will dictate the proton decay rate.
However there are also dimension-6 proton decay diagrams
which do not involve the chirality-flipping mass mixing,
which involve the exchange of $D^c$ or $\bar{D^c}$ only.
Physically the dimension-5 operators correspond to spin-1/2 exchange,
while the dimension-6 operators correspond to spin-0 exchange.

Let us first assume the existence of a symmetry which
prevents the appearance of the term $F{F}D_6 $. However,
this is not enough to prevent proton decay since
$\bar{F}\bar{F}D_6$ contains two operators which are combined
to a proton decay Feynman graph in Fig.1. 
However we observe that if in addition one of the 
colour triplets $D^c$ or $\bar{D^c}$ were to remain heavy,
then fast proton decay would be avoided at least at the
dimension-5 level. How could this be achieved?
The crucial observation is that the mass states in (\ref{Dmass})
are formed between $\bar{d}^c_HD^c$ and 
$d_H^c \bar{D^c}$.
Thus if we extend the symmetry to forbid the term $\bar{H}\bar{H}D_6$
then there is only one allowed mass term for the two triplet pairs,
namely  $<\tilde{\bar{\nu}}_H>\bar{d}^c_HD^c$. When supersymmetry
breaking takes place,  the
scalar part of the other pair receives a $(mass)^2\sim m_{soft}^2$ 
i.e., of the order of the supersymmetry breaking scale.
Thus fast proton decay is avoided, and a phenomenologically
interesting light colour triplet $\bar{D^c}$ and 
colour anti-triplet $d^c_H$ 
pair occur in the spectrum.  

Clearly, the general requirement is that
one of the two operators $F{F}D_6 $ or $\bar{F}\bar{F}D_6$ is forbidden
and in addition one of the operators $\bar{H}\bar{H}D_6$ or
${H}{H}D_6$ is forbidden by the unspecified symmetry.
What is the origin of such a symmetry?
A situation typical of string constructions is to have a
gauge group $G_{PS}\times U(1)$ where
the various matter and higgs multiplets carry  charges $Q_i'$ under
the $U(1)$ symmetry.  {}As an  example, if we assume the charges  
$Q_F'=Q_H'=-3 Q_{\bar F}'=-3 Q_{\bar H}'= -3/2$ and 
$Q_h'=Q_{\phi_i}'=-Q_{D_6}'=-Q_{\varphi}'/2= 1$,  ban the
terms $HH D_6 , FF D_6$ as well as the mixings $\varphi ( H\bar H +
D_6D_6).$ Similarly, making a different choice of the $Q_i'$s, we may
ban the terms $\bar H\bar H D_6 ,\bar F\bar F D_6$.

${\cal D})$. Babu et al \cite{bkmw}, observed 
that there are new kinds of colour
particles which can distinguish between $p e^+$ and
$p e^-$ modes at HERA. In particular in addition to the
$D^c, \bar{D^c}$ fields considered above they also suggest the
following SM representations.
\beq
{\cal U} =  (3,1, \frac 43),\;\; 
{\cal G} =  (3,2, -\frac 76)\;\; ,
\eeq
These states have exotic charges. Both of them can be accommodated in 
representations of our gauge symmetry group. Indeed, assume 
the decomposition of the representation $\Sigma = (15,2,2)$.
\beq
\Sigma = {\cal G}+\bar{{\cal G}} +{\cal U}+\bar{{\cal U}}
+ Q' + \bar{Q}' + h' +\bar{h}' + (8,2)_{\frac 12}+ (8,2)_{-\frac 12}.
\label{sigma}
\eeq
However, if we stick to string motivated scenarios at $k=1$ level\cite{kd},
these representations are not possible since they only arise in the adjoint    
of $SU(4)$. Of course, at $k=2$  Kac-Mody level they are possible.
As an alternative, we may consider that such states may arise as bound
states of smaller representations which bind together due to
their properties  under a hidden symmetry\cite{dlt},
however, in this letter we will not elaborate this further.
%We only wish to mention that some of the extra matter type
%multiplets appearing in the decomposition (\ref{sigma}) may become
%massive by introducing  new irreps. For example, a pair of fourplets
% $H_L= (4,2,1)$,$\bar{H_R}= (\bar{4},1,2}$. 
%Then the coupling $\bar{H}_RH_L\Sigma$ 
%creates mass terms for $Q_H \bar{Q}'$ and $\ell_H\bar{h}'$.

${\cal E})$
We now briefly discuss the problem of unification.
It is known that the particle content of the MSSM  allows the
three gauge couplings to attain a common value at a high scale,
of the order $M_U\sim 10^{16}GeV$.  The introduction of
massless states beyond those of the minimal spectrum change drastically
the evolution of the gauge couplings. Thus, if we assume the
existence  of new (types of) quarks remaining massless down to 
the weak scale, in order that the idea of unification remains intact at
some high scale additional
contibutions to the beta functions are needed to compensate  
for the leptoquark pair and yield a correct prediction for
the weak mixing angle. 
It is clear that if a colour triplet plus anti-triplet pair 
remains in the massless spectum this will alter both the unification
scale and $\sin^2\theta_W$. In the context
of the present model, we desire gauge unification at the string scale,
rather than at $M_U$, so some modification to the spectrum is required
in any case. A complete exhaustive study is clearly required
to determine all possible solutions to the unification question in this model.

Perhaps the simplest unification scenario is one in which we introduce
in addition to the $D^c+\bar{d}^c_H$ or $\bar{D^c}+{d}^c_H$ pair
a further pair of Higgs doublets at low energies contained in an 
extra $h'=(1,2,\bar{2})$ (or perhaps $(1,\bar{2},2)$ )
representation of the Pati-Salam group.
Then the low energy spectrum contains 
an extra two Higgs doublets $h'_1,h'_2$ giving
four Higgs doublets in total. This would clearly have important implications
for the electroweak symmetry breaking sector, which would be interesting
to explore. The low energy spectrum would then contain in addition
to the MSSM spectrum, extra states with the quantum
numbers of an $SU(5)$ $5+\bar{5}$ vector representation, 
and gauge unification is achieved in the usual way at around $10^{16}$ GeV.
If this scale is identified with the Pati-Salam breaking scale
$M_{PS}$ then it is possible to maintain the equality of the coupling
constants right up to the string scale by the addition of suitable
extra heavy Pati-Salam representations \cite{K}.

To summarise, motivated by a possible leptoquark
interpretation of the HERA events, we have discussed the prospects
for including additional light colour triplets and anti-triplets
in the low energy spectrum of unified models.
We have proposed a specific string-inspired Pati-Salam model which
contains a mechanism for allowing low energy states 
which have leptoquark couplings, without inducing excessive proton decay,
and outlined the kind of string symmetries which will forbid the
correct combination of operators to allow this.
There are two possible scenarios:

(1) $HHD_6$ allowed ($\bar{H}\bar{H}D_6$ forbidden) leading to
a light $\bar{D^c}+{d}^c_H$,

(2) $\bar{H}\bar{H}D_6$ allowed ( $HHD_6$ forbidden) leading to a light
$D^c+\bar{d}^c_H$.

In each case unification may be achieved with four
low energy Higgs doublets, and in each case the proton decay
constraint allows one of
two possible options for the leptoquark couplings:

(A) $FFD_6$ allowed ($\bar{F}\bar{F}D_6$ forbidden )
with couplings $\lambda_{9}qq\bar{D^c} +\lambda_{11} q\ell D^c$,

(B) $\bar{F}\bar{F}D_6$ allowed (${F}{F}D_6$ forbidden )
with couplings
$\lambda_{10}u^cd^cD^c + 
\lambda_{12}u^ce^c\bar{D^c}+ \lambda_{13}d^c \bar{D^c}\nu^c $.

Possibility (1B) involves a light leptoquark coupling
$\lambda_{12}u^ce^c\bar{D^c}$, while possibility (2A) 
involves a light leptoquark coupling $\lambda_{11} q\ell D^c$.
Clearly future HERA runs with electron/positron polarisers
will decide which of these two operators is the relevant one.
So far our discussion has been independent of family indices.
Assuming HERA confirms the anomaly, and distinguishes between the
two operators, then the task of theory will be to construct a complete
theory of flavour along these lines. The main point of the present paper
is to show that light leptoquarks may be elegantly
incorporated into string-inspired supersymmetric unified models.

{\it G. K. L. would like to acknowledge stimulating discussions
 with J. Rizos during the initial stages of this work.}

\end{document}